\documentclass[11pt]{article}

\usepackage[final]{acl}

\usepackage{times}
\usepackage{tikz}
\usepackage{amsmath} 
\usepackage{multirow}
\usepackage{amssymb}
\usepackage[most]{tcolorbox}
\usepackage{lipsum} 
\usepackage{latexsym}

\usepackage{comment}
\usepackage{algorithm}
\usepackage{algorithmicx}
\usepackage{algpseudocode}

\usepackage[T1]{fontenc}

\usepackage[utf8]{inputenc}

\usepackage{microtype}

\usepackage{inconsolata}

\newcommand{\blackcircled}[1]{%
  \tikz[baseline=(char.base)]{
    \node[shape=circle,draw,fill=black,inner sep=1.2pt] (char) {\textcolor{white}{\scriptsize #1}};
  }%
}
\usepackage{graphicx}

\newcommand{\stitle}[1]{{{\bf #1}}}

\newcommand{\llmrank}{\textsc{Realm}}
\title{\llmrank: Recursive Relevance Modeling for LLM-based Document Re-Ranking\thanks{{\llmrank} is integrated in: \url{https://ipapers.ai/}.}}

\author{
\textbf{Pinhuan Wang}\textsuperscript{1} \quad
\textbf{Zhiqiu Xia}\textsuperscript{1} \quad
\textbf{Chunhua Liao}\textsuperscript{2} \quad
\textbf{Feiyi Wang}\textsuperscript{3} \quad
\textbf{Hang Liu}\textsuperscript{1} \\
\textsuperscript{1}Rutgers, The State University of New Jersey \quad\\
\textsuperscript{2}Lawrence Livermore National Laboratory \quad
\textsuperscript{3}Oak Ridge National Laboratory \\
\texttt{\{pinhuan.wang, hang.liu\}@rutgers.edu}
}

\begin{document}
\maketitle
\begin{abstract}
Large Language Models (LLMs) have shown strong capabilities in document re-ranking, a key component in modern Information Retrieval (IR) systems. However, existing LLM-based approaches face notable limitations, including ranking uncertainty, unstable top-$k$ recovery, and high token cost due to token-intensive prompting. To effectively address these limitations, we propose \llmrank, an uncertainty-aware re-ranking framework that models LLM-derived relevance as Gaussian distributions and refines them through recursive Bayesian updates. By explicitly capturing uncertainty and minimizing redundant queries, {\llmrank} achieves better rankings more efficiently. Experimental results demonstrate that our {\llmrank} surpasses state-of-the-art re-rankers while significantly reducing token usage and latency, improving NDCG@10 by $0.7-11.9$ and simultaneously reducing the number of LLM inferences by $23.4-84.4\%$, promoting it as the next-generation re-ranker for modern IR systems. 
\end{abstract}

\section{Introduction}
Document re-ranking is a key component in modern IR systems~\cite{zhu2023large}. Given a user query, retrieval systems typically begin with a fast but coarse retrieval stage that returns a broad set of potentially relevant documents. However, these initial results are often noisy or only loosely related to the query. Re-ranking addresses this issue by applying a more accurate, context-aware scoring model to refine the order of the candidates and place the most relevant documents at the top~\cite{nogueira2019passage}.
For example, in academic paper searching, an initial retrieval step may return a large number of documents that match surface-level keywords, while missing more in-depth relevance analysis. Without re-ranking, the most important references could be placed at the bottom, potentially leading researchers to miss those key published findings in their research journey. 
In a nutshell, re-ranking is crucial to ensure that high-quality, contextually appropriate documents are selected as input for subsequent applications~\cite{lewis2020retrieval}. 


LLMs are redefining document re-ranking by enabling deep semantic and contextual understanding that traditional methods fundamentally lack. Traditional re-rankers, such as those based on BM25 scores~\cite{robertson2009probabilistic} or learning-to-rank models like LambdaMART~\cite{burges2010ranknet}, rely on either simple sparse feature-term overlap, document frequency, or hand-crafted heuristics, which often fail in capturing nuanced relevance, especially in complex or fine-grained queries. In contrast, LLM-based re-rankers treat the query and candidate documents as joint inputs, allowing for more in-depth relevance estimation grounded in deep semantic and contextual comprehension. Recent studies have shown that LLMs, when applied as cross-encoders or guided with task-specific prompting, consistently outperform classical re-rankers across benchmarks~\cite{sun-etal-2023-chatgpt}. These advancements suggest that LLMs are not just an incremental improvement but a paradigm shift toward unifying retrieval and comprehension within a single, adaptable framework.

However, LLM-based document re-ranking faces a three-pronged challenge:  
(i) \textbf{ranking uncertainty}, stemming from the inherent stochastic nature of LLMs (see Section~\ref{subsec:uncertain});  
(ii) \textbf{unstable top-$k$ recovery}, where minor input variations can substantially disrupt document rankings (see Figure~\ref{fig:sensitivity}); and  
(iii) \textbf{high token costs}, due to the need of complex prompting strategy (see Table~\ref{tab:overall}).

Recent research endeavors have fallen short in effectively addressing all three challenges: 
\textbf{Pointwise methods}~\cite{nogueira-etal-2020-document} are efficient and parallelizable, as they assess each document independently. Some variants~\cite{zhuang-etal-2023-open} further leverage generation likelihood as a relevance score. However, pointwise approaches fail to model interactions among candidates, making them less effective at resolving uncertainty or producing globally consistent top-$k$ rankings.
\textbf{Listwise methods}~\cite{sun-etal-2023-chatgpt} enable joint evaluation of multiple candidates in a single query, which helps mitigate ranking inconsistency. Approaches like TourRank~\cite{chen2024tourrank} adopt tournament-style aggregation to extend listwise scoring. Despite these benefits, listwise methods still suffer from context length constraints and positional bias~\cite{liu-etal-2024-lost}, especially for long candidate sets.
\textbf{Pairwise methods}~\cite{qin-etal-2024-large} improve local comparison quality by directly modeling relative preferences between document pairs. Advanced systems like PRP-Graph~\cite{luo-etal-2024-prp} further exploit graph structures to aggregate pairwise signals. Nevertheless, the repeated comparisons lead to high token usage and substantial inference latency.
\textbf{Setwise methods}~\cite{zhuang2024setwise,podolak2025beyond} improve efficiency by evaluating small subsets at a time, but discard fine-grained preference information, such as full relevance logits—thereby under-utilizing the model’s capacity.

To effectively address the three-pronged challenge, this paper proposes \llmrank, an uncertainty-aware re-ranking framework that combines relevance estimation with a recursive refinement process. 
{\llmrank} explicitly models uncertainty, improves top-$k$ stability, and reduces inference costs.
Our contributions are as follows:
\begin{itemize}
\item \textbf{Uncertainty-Aware Relevance Modeling.} We model each document’s relevance as a Gaussian distribution, capturing both the estimated score and uncertainty to support robust re-ranking under the inherent stochastic nature of contemporary LLMs.
\item \textbf{Recursive Refinement Framework.} We introduce a recursive framework that compares pivot documents with subsets and refines relevance distributions through Bayesian updates, enhancing ranking stability.
\item \textbf{Pivot-Centric Optimizations.} We optimize efficiency by selecting high-confidence pivots, aggregating updates via uncertainty-aware averaging, and applying pivot adjustment to ensure effective workload reduction.

\end{itemize}

Experiments show that {\llmrank} outperforms state-of-the-art re-ranking methods while substantially reducing token usage and improving stability, making it suitable for real-world retrieval systems.

\section{Related Work \& Preliminary}

\subsection{Related Work}

Zero-shot document re-ranking with LLMs is typically grounded in four fundamental prompting strategies: pointwise~\cite{nogueira-etal-2020-document}, pairwise~\cite{qin-etal-2024-large}, listwise~\cite{sun-etal-2023-chatgpt}, and setwise~\cite{zhuang2024setwise}. In this section, we compare {\llmrank} with these papers.

\stitle{Pointwise methods.}
Pointwise methods prompt LLMs to assess the relevance of each document independently with respect to a given query, typically by generating a relevance score or extracting the score from the output logits~\cite{nogueira-etal-2020-document}. Several variants exist. For example, Query Generation~\cite{zhuang-etal-2023-open} estimates query-document compatibility by computing the likelihood of the query given a passage. 
While these approaches are token efficient and scalable, they struggle to capture comparative relevance across candidates, which is well preserved in {\llmrank}.

\stitle{Listwise methods.}
With the continued expansion of LLM capacity and input window size, listwise ranking, where the model receives a group of candidate documents and directly outputs their relative ordering, has become increasingly feasible. By supporting joint reasoning over multiple candidates within a single inference, this paradigm has motivated a series of methods aimed at better leveraging LLM capabilities for document re-ranking.

RankGPT~\cite{sun-etal-2023-chatgpt} and LRL~\cite{ma2023zero} adopt a sliding-window listwise re-ranking strategy, comparing a subset of candidates at each step, retaining the most relevant ones, and forwarding the rest for subsequent comparisons. TourRank~\cite{chen2024tourrank} draws inspiration from sports tournaments, treating each subset as a group match and aggregating results through a point-based system. ListT5~\cite{yoon-etal-2024-listt5} follows a similar tournament-style design, effectively implementing an $m$-ary heap traversal over listwise scoring primitives to recover the top-$k$ results.

Despite their effectiveness, these methods face inherent limitations. Current LLMs are still restricted by finite context lengths and remain sensitive to positional biases~\cite{liu-etal-2024-lost}, which hinders their ability to process long candidate lists holistically and maintain consistency across multiple comparisons.
In contrast, {\llmrank} avoids full-list comparisons by decomposing the ranking process into a sequence of setwise updates. This approach enables consistent top-$k$ selection without suffering from context-length limitations or positional bias, while still leveraging the LLM's capacity to reason over small candidate sets.

\stitle{Pairwise methods.}
Pairwise prompting (PRP) \cite{qin-etal-2024-large} was introduced to overcome the limitations of pointwise and listwise ranking by prompting the model to compare two candidates at a time and choose the more relevant one. To extend this into a full ranking, the authors implemented a multi-round bubble sort using overlapping comparisons to extract the top-$k$ candidates.
PRP-Graph~\cite{luo-etal-2024-prp} further generalizes this idea by constructing a weighted comparison graph and applying a PageRank-style aggregation to derive a global ranking.

While pairwise prompting yields accurate comparisons, it incurs high token costs due to repeated queries.
In contrast, our method reduces the number of LLM calls by performing aggregation over setwise comparison, achieving greater efficiency without sacrificing ranking quality.

\stitle{Setwise methods.} Setwise prompting~\cite{zhuang2024setwise} was introduced as a refined variant of listwise prompting, leveraging model output logits to select the top-$k$ documents within a group. This strategy was extended by integrating it with classic sorting algorithms such as bubble sort and heap sort.
Setwise Insertion~\cite{podolak2025beyond} further advanced this line of work by incorporating the initial document ranking as prior knowledge, thereby improving ranking efficiency.

While drawing inspiration from setwise prompting, {\llmrank} preserves the full comparative information encoded in logits and performs uncertainty-aware updates through probabilistic aggregation, enabling more robust relevance estimation.

\stitle{Other directions in LLM for re-ranking.}
%
\textit{(i)} Training strategies for LLM-based re-rankers.
RankT5~\cite{zhuang2023rankt5} adopts pairwise and listwise training objectives for T5, while ChainRank-DPO~\cite{liu2024chainrank} enhances ranking consistency using CoT-style supervision with DPO. 
Rank-R1~\cite{zhuang2025rank} introduces reinforcement learning with limited supervision to promote reasoning over queries and documents.
RankGPT~\cite{sun-etal-2023-chatgpt} and RankVicuna~\cite{pradeep2023rankvicuna} distill ChatGPT/GPT-3.5 into smaller models via pairwise and listwise losses. ListT5~\cite{yoon-etal-2024-listt5} uses Fusion-in-Decoder for listwise inference, and TSARankLLM~\cite{zhang-etal-2024-two} adopts a two-stage pretraining and fine-tuning strategy.

%
%
%
\textit{(ii)} Hybrid architectures.
Hybrid methods restructure the inference process by combining ranking components or decomposing tasks,  e.g., EcoRank~\cite{rashid-etal-2024-ecorank}, RankFlow~\cite{jin2025rankflow}, and APEER~\cite{jin2025apeer}.
In parallel, Permutation Self-Consistency~\cite{tang-etal-2024-found} aggregates multiple permutations to reduce positional bias, and LLM-RankFusion~\cite{zeng2024llm} improves robustness via calibration and fusion-based aggregation.

\subsection{LLM Uncertainty \& Bayesian Rating System}
\label{subsec:uncertain}

\stitle{LLM uncertainty.}
LLMs exhibit two forms of uncertainty: aleatoric, which stems from inherent data noise, and epistemic, which arises due to limited training coverage~\cite{kendall2017uncertainties}. While aleatoric uncertainty is largely irreducible, epistemic uncertainty can be mitigated by introducing additional informative signals during inference.

In this context, studies have shown that in multiple-choice settings, LLM output logits are often well-calibrated—i.e., their relative magnitudes reliably reflect the model's confidence~\cite{kadavath2022language, si-etal-2023-measuring}. This calibration enables softmax-normalized logits to serve as meaningful probability estimates, supporting downstream applications such as uncertainty estimation and probabilistic relevance modeling.

\begin{figure*}[t]
  \centering
  \includegraphics[width=\linewidth]{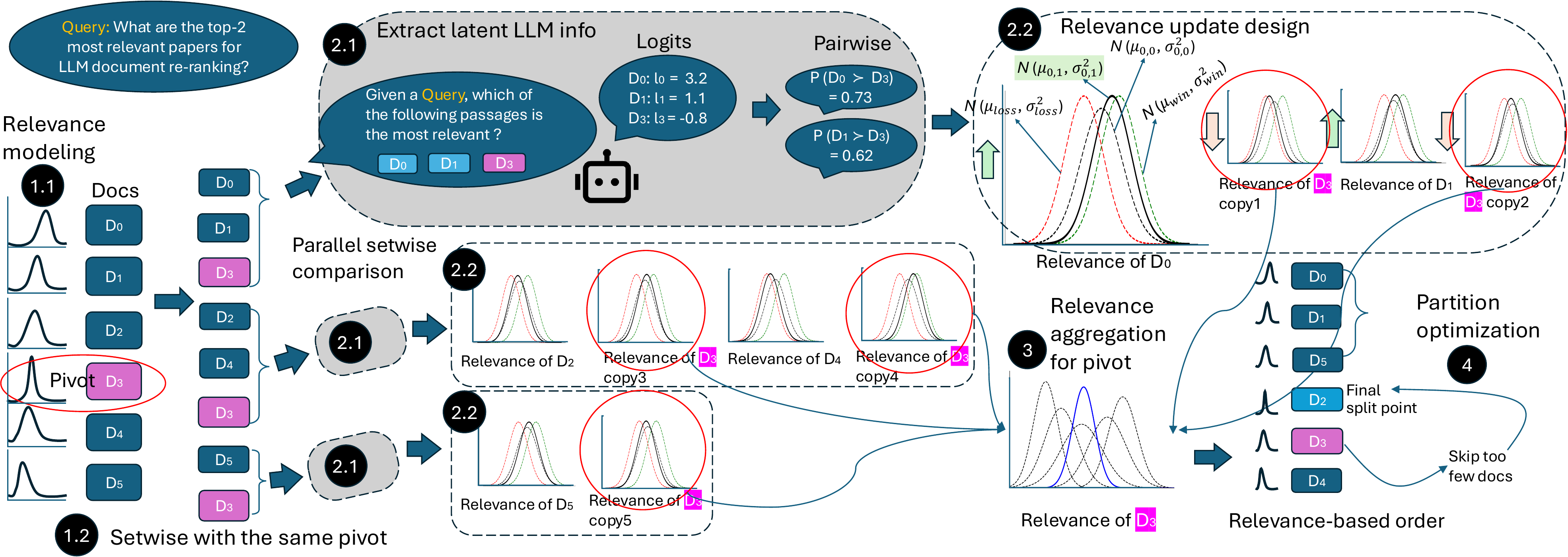}
  \caption{Workflow of our recursive relevance modeling framework for LLM-based document re-ranking.
  }
  \label{fig:dis}
\end{figure*}

\stitle{Bayesian rating systems} offer principled probabilistic frameworks for estimating latent skill levels or quality scores based on observed outcomes from comparisons or matches. Grounded in Bayesian inference, these systems iteratively update skill estimates by combining prior beliefs with new evidence. Key advantages include explicit uncertainty modeling, incremental update capabilities, and robustness to noisy or incomplete data. Two widely adopted instances are the Elo rating system~\cite{elo} and its more expressive successor, TrueSkill~\cite{herbrich2006trueskill}, which extend the rating process to handle more complex scenarios, which we detail in the Appendix~\ref{app:rating}.

\section{Methodology}

\subsection{{\llmrank} Framework}

Figure~\ref{fig:dis} illustrates the workflow of our relevance modeling framework for LLM-based document re-ranking. As an illustrative example, the task is to retrieve the top-$2$ most relevant documents on the topic of LLM-based re-ranking. Given a set of documents, we model each document’s relevance as a Gaussian distribution $\mathcal{N}(\mu, \sigma^2)$ (step~\blackcircled{1.1}).

We begin by selecting a pivot $D_3$ to put the documents into multiple subsets. For each subset, we conduct a setwise comparison involving the same pivot, as shown in step~\blackcircled{2.1} in Figure~\ref{fig:dis}. Then, as depicted in step~\blackcircled{2.2}, we adopt a Bayesian update to refine the relevance distributions. Further details are presented in $\S$~\ref{subsec:bayesian}.
Further, as shown in step~\blackcircled{ 3 }, we introduce a mechanism to update document $D_3$'s relevance model. 

A naive subsequent design of {\llmrank} would directly rely on the relevance distributions of the first iteration to order all the documents. Subsequently, we can select the top-$k$ most relevant documents as the final result. Particularly, we can derive the relevance score of each document using a distribution-based rule \(\mu - k\sigma\), where \(k\) is a constant controlling conservativeness, with higher values penalizing uncertainty more heavily. 
However, this design could potentially rely on relevance distributions that are very unstable, as a single round of comparison might fail to effectively curb the uncertainty of the relevance distributions (see Table~\ref{tab:ablation}). 

Consequently, we introduce a recursive design, that is, we compare each document against the pivot document. If a document is closer to the query than the pivot, we keep it for the next round of calculation. Otherwise, we filter out that document. 
%
%
Moreover, directly relying on the pivot $D_3$ to filter out unpromising documents would yield unstable workload reduction. We thus design an effective workload reduction mechanism to cope with this concern, which could derive $D_4$ as the final split point (\blackcircled{ 4 }). Of note, we also design a strategy to select the document with the highest confidence as the pivot (step~\blackcircled{1.2}).


\subsection{\llmrank's Relevance Modeling Scheme}
\label{subsec:bayesian}

\textbf{Modeling relevance as a normal distribution.}
In {\llmrank}, relevance judgment derived from LLMs is inherently noisy due to their contextual sensitivity, response stochasticity, and inherent biases~\cite{dai2024bias}.
To capture this uncertainty, following TrueSkill~\cite{herbrich2006trueskill}, a framework originally designed for competitive player rating, we model the relevance of each document to a given query as a Gaussian distribution, which is denoted as $\mathcal{N}(\mu, \sigma^2)$, where the mean $\mu$ represents the estimated relevance score and the variance $\sigma^2$ quantifies the uncertainty of this estimate. 

As illustrated in step~\blackcircled{1.1} of Figure~\ref{fig:dis}, each document is initially associated with an \textit{initial relevance distribution} (i.e., the dark blue squiggle lines beside each document $D_i$). We model the relevance as a Gaussian distribution with fixed standard deviation~\(\sigma_0\), reflecting a uniform level of uncertainty across all unobserved documents. The initial mean~\(\mu_0\) is set based on the document’s retrieval score if available (e.g., from a pre-LLM retrieval pipeline such as embedding search); otherwise, a shared default value is assigned.


\stitle{Extracting latent LLM information for relevance model update.}
To refine these initial relevance distributions, \llmrank\ extracts latent information from LLMs during setwise comparison. 


In setwise comparison, LLM is prompted to select the most relevant document from a small group (of size $m$) of candidate documents. Our example query is "Given a query, which of the following passages is the most relevant?". Internally, LLM assigns scalar logits to each option; we extract these as a score vector~\(\{\ell_{i} \colon D_{i} \in \text{set}\}\), where each \(\ell_i\) corresponds to the document~\(D_i\). For instance, in step~\blackcircled{2.1} of Figure~\ref{fig:dis}, the logits assigned to $(D_0, D_1, D_3)$ are $3.2$, $1.1$, and $-0.8$, respectively. Of note, for closed-source LLMs, we can prompt them to output the confidence values and use that as the scores~\cite{xia2025survey}. 

Rather than relying on these scores for direct selection or ranking~\cite{zhuang2024setwise}, \llmrank\ interprets their differences as pairwise preference probabilities. Specifically, the probability that document $D_i$ is preferred over $D_j$ is computed as:
\[
P(D_i \succ D_j) = \sigma\left(\frac{\ell_i - \ell_j}{T}\right),
\]
where $\sigma$ is the sigmoid function and $T$ is a temperature parameter. This formulation transforms a single multi-document comparison into a set of $\binom{m}{2}$ pairwise probability updates. 


To ensure consistency in aggregation, \llmrank\ adopts a pivot-based strategy. One document in the prompt is designated as the pivot, and preference probabilities are computed between the pivot and the others. As illustrated in step~\blackcircled{2.1} of Figure~\ref{fig:dis}, when $D_3$ is used as the pivot alongside $D_0$ and $D_1$, we extract the probabilities $P(D_0 \succ D_3)$ and $P(D_1 \succ D_3)$. These preference probabilities are then used to update the relevance distributions of both the pivot and its comparators as follows.

\stitle{Relevance update design.} The extracted pairwise preference probabilities, such as $P(D_0 \succ D_3)$ and $P(D_1 \succ D_3)$ from step~\blackcircled{2.1}, are then used to update each document’s relevance distribution. 

Rather than treating each update as a deterministic outcome (win, loss, or draw) as in the original TrueSkill, we leverage the preference probability to capture richer information in relevance updates.
Specifically, we interpolate the influence of the win and loss outcomes based on the predicted probability, i.e., \(P(D_i \succ D_j)\).

As illustrated in step~\blackcircled{2.2}, the model predicts \(P(D_0 \succ D_3) = 0.73\). In this case, the relevance distribution of \(D_0\), i.e., $\mathcal{N}(\mu_{0,0}, \sigma_{0,0}^2)$, is updated to $\mathcal{N}(\mu_{0,1}, \sigma_{0,1}^2)$ by interpolating 
between the two TrueSkill-updated distributions: a win $\mathcal{N}(\mu_{win}, \sigma_{win}^2)$ (green dashed curve) and a loss $\mathcal{N}(\mu_{loss}, \sigma_{loss}^2)$ (red dashed curve) against \(D_3\), weighted by the predicted probability.

The resulting distribution for \(D_0\) thus shifts toward the win-specific distribution while incorporating uncertainty, effectively reflecting both the model's directional preference and its confidence.

Formally, let the distribution of document~\(D_i\) be \(\mathcal{N}(\mu_{i,0}, \sigma_{i,0}^2)\), with natural parameters defined as the {precision} \(\lambda_{i,0} = 1/\sigma_{i,0}^2\) and the {precision-adjusted mean} \(\tau_{i,0} = \mu_{i,0}/\sigma_{i,0}^2\).  
We compute two updated distributions for \(D_i\) by applying the {standard TrueSkill update rules} for a 1v1 match against \(D_j \sim \mathcal{N}(\mu_{j,0}, \sigma_{j,0}^2)\), as follows:

\begin{itemize}
  \item \(\mathcal{N}\left(\mu_{\text{win}}, \sigma^2_{\text{win}}\right)\), assuming \(D_i\) wins over \(D_j\), 
  \item \(\mathcal{N}\left(\mu_{\text{loss}}, \sigma^2_{\text{loss}}\right)\), assuming \(D_i\) loses to \(D_j\).
\end{itemize}

Let \(\tau_{\text{win}} = \mu_{\text{win}} / \sigma^2_{\text{win}}\) and \(\lambda_{\text{win}} = 1 / \sigma^2_{\text{win}}\), and similarly for the loss outcome.  
Given a win probability \(p = P(D_i \succ D_j)\), we apply a fractional update~\cite{minka2004power} in the natural parameter space by combining the additive changes from the win/loss outcomes:
\vspace{-.15in}
\[
\lambda_{i,1} = \lambda_{i,0} + p \cdot (\lambda_{\text{win}} - \lambda_{i,0}) + (1 - p) \cdot (\lambda_{\text{loss}} - \lambda_{i,0}),
\]
\[
\tau_{i,1} = \tau_{i,0} + p \cdot (\tau_{\text{win}} - \tau_{i,0}) + (1 - p) \cdot (\tau_{\text{loss}} - \tau_{i,0}),
\]

The resulting distribution of \(D_i\) is again a Gaussian, given by:
\[
\mathcal{N}\left(\mu_{i,1}, \sigma^2_{i,1}\right), ~\text{where}~
\mu_{i,1} = \frac{\tau_{i,1}}{\lambda_{i,1}},~
\sigma^2_{i,1} = \frac{1}{\lambda_{i,1}}.
\]


This relevance update enables our model to integrate both the direction and confidence of each comparison. Further details are provided in Appendix~\ref{app:update}. As shown in step~\blackcircled{2.2} of Figure~\ref{fig:dis}, \textit{the updated distribution becomes narrower, indicating increased certainty and shifts toward the more likely outcome}, resulting in a more accurate and confident relevance distribution.

\subsection{Pivot-Centric Optimizations
}\label{subsec:quicksort}


\stitle{Pivot aggregation.}
The previous pivot-based strategy enables the extraction of useful pairwise comparisons centered on each pivot. To integrate global relevance signals for a given pivot, we introduce \textit{pivot aggregation} that consolidates its comparison outcomes across all subsets.


After the current round, we combine the relevance models of various copies for the pivot via an uncertainty-aware averaging:
\[
\begin{aligned}
  \tau            &= \sum_{i=1}^{c}\sigma_i^{-2},\\
  \mu_{\mathrm{agg}} &= \frac{\sum_{i=1}^{c}\mu_i\,\sigma_i^{-2}}{\tau},\\
  \sigma_{\mathrm{agg}} &= \left(\frac{\tau}{n}\right)^{-1/2}.
\end{aligned} \label{eq:soft}
\]
Here, $\tau$ denotes the total precision accumulated from $c$ shadow comparisons, and the averaging yields an aggregated distribution that favors more confident estimation while reducing overall uncertainty.
As shown in step~\blackcircled{ 3 } of Figure~\ref{fig:dis}, the pivot $D_3$ is replicated into five copies and updated independently through soft comparisons. The final distribution of $D_3$ is then obtained by aggregating these updates, resulting in a more stable and unbiased estimation.

\stitle{Pivot selection.}
To determine the global pivot at each recursive step, we select the document with the lowest estimated standard deviation~$\sigma$ from the current candidate pool. Intuitively, a lower $\sigma$ indicates higher confidence in the document’s relevance estimate. Using such a document as the pivot improves the stability of the partitioning process. 
Initially, the $\sigma$ may be the same across documents. In this case, we select the document with the median estimated relevance score $\mu$ as the pivot, to avoid pruning too many or too few candidates.
As illustrated in step~\blackcircled{1.2} of Figure~\ref{fig:dis}, document~$D_3$ is selected as the global pivot due to its lowest uncertainty. The remaining documents are then grouped into prompts: $\{D_0, D_1\}$, $\{D_2, D_4\}$, and $\{D_5\}$, ensuring that each non-pivot document is compared once against $D_3$. This structure enables us to construct a globally consistent relevance preference centered on a high-confidence document.

\stitle{Pivot adjustment for effective reduction.}
%
%
%
%
Considering that pivot might not be able to effectively reduce the documents, we interpolate the pivot with the interval midpoint:
\[
  i^* = \lambda\, r_p + (1-\lambda)\,\frac{l + r}{2}, \quad \lambda \in [0,1],
\]
where \(l\) and \(r\) are the current interval bounds. Using \(i^*\) as the split index helps avoid unbalanced partitions. Setting \(\lambda < 1\) guarantees recursion depth remains bounded by \(O(\log n)\) and total LLM comparisons scale linearly with \(n\). While the pivot still guides comparisons, the softened partition is used only to improve efficiency and does not change the underlying ranking based on the pivot.
As illustrated in step~\blackcircled{ 4 } of Figure~\ref{fig:dis}, only documents ranked above the split point (e.g., $D_3$) are retained for the next iteration of refinement. This iteration continues until the desired top-$k$ set is extracted.

\begin{table*}
  \centering
    \resizebox{1\linewidth}{!}{ 
  \begin{tabular}{c|c|ccccc|ccccc}
    \hline
    \multirow{2}{*}{\textbf{LLM}}&\multirow{2}{*}{\textbf{Method}} & \multicolumn{5}{|c|}{\textbf{TREC DL 2019}}& \multicolumn{5}{|c}{\textbf{TREC DL 2020}}\\
    \cline{3-12}
    &&\textbf{N@10}&\textbf{\#Inf.}&\textbf{P. tks.}&\textbf{G. tks.}&\textbf{Lat.(s)}&\textbf{N@10}&\textbf{\#Inf.}&\textbf{P. tks.}&\textbf{G. tks.}&\textbf{Lat.(s)}\\\hline
     NA&BM25& 50.6  &- &- &-&- &48.0 &-&-&-&- \\\hline
     \hline
     \multirow{5}{*}{\rotatebox{90}{Flan-T5-Large}}
     &TourRank&48.2&130.0&95271.4& 1507.2&56.9&40.7&130.0&95341.8&1524.5&  57.1\\\cline{2-12}
     &PRP-Graph&65.8&492.7&221781.9&-&43.3&61.8&492.5&224605.5&-&42.4\\\cline{2-12}
     &Setwise-Heapsort&66.9&125.3&40449.6&626.5&8.8&61.8&124.2&40357.4&621.0&8.7 \\\cline{2-12}
      &Setwise-Insertion&66.9&92.5&29913.1&93.4&4.5&62.5&91.3&29757.7&93.8&4.4\\\cline{2-12}      
     &{\llmrank}&{\textbf{\underline{67.0}}}&\textbf{\underline{79.0}}&\textbf{\underline{25165.7}}&-&\textbf{\underline{3.9}}&\textbf{\underline{63.0}}&\textbf{\underline{74.6}}&\textbf{\underline{23584.9}}&-&\textbf{\underline{3.6}}\\\hline
     \hline
     \multirow{5}{*}{\rotatebox{90}{Flan-T5-XL}}
    &TourRank&64.3&130.0&95257.0&2719.2&96.8&59.7&130.0&95277.6&2791.0&100.6\\\cline{2-12}
      &PRP-Graph&67.6&492.6&212884.5&-&43.0&66.1&492.5&216071.3&-&43.1\\\cline{2-12}
     &Setwise-Heapsort &69.2&129.5&41665.7&647.7&10.1&67.8&127.8&41569.1&639.1&9.6\\\cline{2-12}
     &Setwise-Insertion&69.0&106.0&34732.7&100.7&5.3&67.0&105.3&34400.7&99.5&5.1\\\cline{2-12}
     &{\llmrank}&\textbf{\underline{70.5}}&\textbf{\underline{80.4}}& \textbf{\underline{25823.0}}&-&\textbf{\underline{4.0}}&\textbf{\underline{68.4}}&\textbf{\underline{75.6}}&\textbf{\underline{24033.6}}&-&\textbf{\underline{3.7}}\\\hline
     \hline
     \multirow{5}{*}{\rotatebox{90}{Flan-T5-XXL}}
      &TourRank&61.9&130.0&95269.3&1610.4&133.6&62.7&130.0&95273.6&1615.6&133.2\\\cline{2-12}
     &PRP-Graph&66.6&492.6& 213536.2&-&73.3&66.1&492.6&216332.4&-&74.4\\\cline{2-12}
     &Setwise-Heapsort &70.6 & 130.1&42078.6&650.5& 15.9&68.8&128.2&41633.7&640.8&15.7\\\cline{2-12}
     &Setwise-Insertion&68.4&104.9&34284.2&99.2&10.6&67.1&100.7&33036.9&100.0&10.2\\\cline{2-12}
    &{\llmrank}&\textbf{\underline{71.2}}&\textbf{\underline{76.5}}& \textbf{\underline{24659.8}}&-&\textbf{\underline{7.5}}&\textbf{\underline{69.1}}&\textbf{\underline{74.1}}& \textbf{\underline{23759.6}}&-&\textbf{\underline{7.3}}\\\hline
    \end{tabular}
    }
    \caption{Evaluation on TREC DL 2019 and TREC DL 2020 datasets: {\llmrank} vs TourRank~\cite{chen2024tourrank}, PRP-Graph~\cite{luo-etal-2024-prp}, Setwise-Heapsort~\cite{zhuang2024setwise}, and Setwise-Insertion~\cite{podolak2025beyond}.}
  \label{tab:overall}
\end{table*}

\section{Experiments}

\subsection{Experimental Setup}

We conduct evaluations on Flan-T5 models~\cite{longpre2023flan} of three sizes—Flan-T5-Large (770M parameters), Flan-T5-XL (3B), and Flan-T5-XXL (11B)-following recent work on LLM-based re-ranking~\cite{qin-etal-2024-large,luo-etal-2024-prp,zhuang2024setwise,podolak2025beyond}.
To assess the generality of {\llmrank}, we additionally evaluate Flan-UL2 (20B)~\cite{tay2022ul2} and LLaMA3 (8B and 70B)~\cite{grattafiori2024llama}. Notably, LLaMA3 follows a decoder-only architecture and differs from Flan-T5 models in both model structure and pretraining objectives.

All experiments are conducted on a server with $512$ GB RAM, two Intel Xeon Silver 4309Y CPUs ($16$ cores), and four A100 GPUs ($80$ GB each). All models are evaluated on a single GPU, except for LLaMA3-70B, which uses all four GPUs.

\stitle{Datasets and metrics.} 
Experiments are conducted on widely used benchmarks: TREC Deep Learning 2019~\cite{craswell2020overview}, 2020~\cite{craswell2021overview}, and the BEIR benchmark~\cite{BEIR}.

All LLM-based methods re-rank the top $100$ documents retrieved by a BM25 first-stage retriever. Adopting prior work's evaluation strategy~\cite{zhuang2024setwise,podolak2025beyond}, we formulate re-ranking as a top-$k$ task, with $k=10$ as the default setting. Effectiveness is measured using the NDCG@10 metric for all datasets. For clarity, all NDCG@10 results are presented as percentages.


\stitle{Baselines.}  
We compare our method with four recent LLM-based re-ranking approaches: TourRank~\cite{chen2024tourrank}, PRP-Graph~\cite{luo-etal-2024-prp}, Setwise-Heapsort~\cite{zhuang2024setwise}, and Setwise-Insertion~\cite{podolak2025beyond}. Implementation details are provided in Appendix~\ref{app:imp}. 

TourRank is a state-of-the-art listwise re-ranking method inspired by sports tournaments. It treats each subset of documents as a ``group match'' and aggregates the results using a point-based system.  
PRP-Graph constructs a global ranking by aggregating local pairwise preferences through a graph-based approach.  
Setwise-Heapsort and Setwise-Insertion utilize setwise prompting to compare multiple candidates jointly in a token-efficient way. The former focuses on computational efficiency using a heap-based sorting strategy, while the latter improves ranking accuracy through a more refined insertion-based sorting mechanism.




\subsection{Overall Evaluation}

Table~\ref{tab:overall} presents a comprehensive comparison of our approach on both the TREC-DL 2019 and 2020 benchmarks. We report NDCG@10 (N@10), Inf. (inference counts, a.k.a., the \# of LLM calls), P. tks. (\#tokens in prompt), G. tks. (\# of generated tokens), and Lat. (latency in seconds).

Compared to TourRank, PRP-Graph, Setwise-Heapsort, and Setwise-Insertion, {\llmrank} achieves consistent improvements in both ranking quality and inference efficiency across all evaluated benchmarks. On average, {\llmrank} outperforms all baselines, improving NDCG@10 by $0.7-11.9$, while simultaneously reducing the number of LLM inferences by $23.4-84.4\%$ and cutting inference latency by $25.0-88.7\%$.
In terms of prompt token usage (P. tks.), {\llmrank} reduces the cost by $25.2-94.8\%$. Furthermore, since it only leverages the logits of the first generated token, the generation token cost (G. tks.) is effectively eliminated.

On the model size dimension, we observe that Flan-T5-XL outperforms Flan-T5-Large, and Flan-T5-XXL further surpasses Flan-T5-XL, aligning with the general trend that larger instruction-tuned models exhibit stronger ranking capabilities. Among all methods, TourRank demonstrates the highest sensitivity to model capacity, as its listwise comparison approach relies heavily on both the model's input context length and its vulnerability to positional biases.

While PRP-Graph and TourRank consume the most tokens, this is mainly due to their reliance on multiple iterative rounds of pairwise or listwise comparisons to accumulate sufficient preference information, resulting in significantly higher total query costs. Setwise-Heapsort and Setwise-Insertion offer a more favorable efficiency-performance trade-off by utilizing structured comparisons with fewer rounds. However, they still underutilize the rich preference information embedded in the LLM’s output, such as the model’s confidence in its ranking decisions.
This leaves room for further enhancement of {\llmrank} by incorporating more principled information aggregation strategies and refined comparison scheduling to fully leverage the LLM's capacity.
\begin{table*}
  \centering
    \resizebox{1\linewidth}{!}{ 
  \begin{tabular}{c|c|ccccc|ccccc}
    \hline
    \multirow{2}{*}{\textbf{LLM}}&\multirow{2}{*}{\textbf{Method}} & \multicolumn{5}{|c|}{\textbf{BEIR Covid}}& \multicolumn{5}{|c}{\textbf{BEIR SciFact}}\\
    \cline{3-12}
    &&\textbf{N@10}&\textbf{\#Inf.}&\textbf{P. tks.}&\textbf{G. tks.}&\textbf{Lat.(s)}&\textbf{N@10}&\textbf{\#Inf.}&\textbf{P. tks.}&\textbf{G. tks.}&\textbf{Lat.(s)}\\\hline
     NA&BM25& 59.5  &- &- &-&- &67.9 &-&-&-&- \\\hline
     \hline
     \multirow{5}{*}{\rotatebox{90}{Flan-T5-Large}}
     &TourRank&44.4&130.0&96946.4& 1670.3&59.1&15.7&130.0&97816.3	&1679.0&	62.2\\\cline{2-12}
     &PRP-Graph&77.2	&492.5	&309267.7&-&53.7	&64.6&	492.0&	315965.2&-&	52.9\\\cline{2-12}
     &Setwise-Heapsort&75.4&	129.6&	58286.4&648.0&8.7&62.0&119.2&	54641.1&596.0&7.9\\\cline{2-12}
      &Setwise-Insertion&74.2&120.3&	53851.7&96.5&5.6&56.2	&148.1&	67053.6	&88.2&6.7\\\cline{2-12}      
     &{\llmrank}&\textbf{\underline{78.4}}&\textbf{\underline{81.2}}&\textbf{\underline{36365.8}}&-&\textbf{\underline{4.3}}&\textbf{\underline{69.3}}&\textbf{\underline{69.8}}&\textbf{\underline{31901.7}}&-&\textbf{\underline{3.7}}\\\hline
      & & \multicolumn{5}{|c|}{\textbf{BEIR DBpedia}}& \multicolumn{5}{|c}{\textbf{BEIR NFCorpus}}\\
    \hline

     NA&BM25& 31.8  &- &- &-&- &32.2 &-&-&-&- \\\hline
     \hline
     \multirow{5}{*}{\rotatebox{90}{Flan-T5-Large}}

&TourRank	&28.8&	130.0&	95231.8&	1466.0&56.8&30.6&	130.0&93211.3&1551.3&43.4     \\\cline{2-12}
&PRP-Graph	&44.0&	491.4&236478.8&-&43.3&33.9&349.2&214268.8	&-	&32.1     \\\cline{2-12}
&Setwise-Heapsort&41.3&124.5&41529.0&622.3&8.1&32.4&87.8	&38856.5&438.3&6.5     \\\cline{2-12}
&Setwise-Insertion	&42.0&109.3&36393.6&93.6&4.6&32.0&	80.3&35314.2&76.4&3.9     \\\cline{2-12}
&{\llmrank}&\textbf{\underline{45.9}}&\textbf{\underline{75.4}}&\textbf{\underline{25848.0}}&-&\textbf{\underline{3.6}}&\textbf{\underline{36.4}}&\textbf{\underline{50.2}}&\textbf{\underline{22222.4}}	&-&	\textbf{\underline{2.7}}\\

     \hline
    
    \end{tabular}
    }
    \caption{Supplementary evaluation on four BEIR datasets Covid, SciFact, DBpedia, and NFCorpus, using Flan-T5-Large model.}
  \label{tab:supple}
\end{table*}

\begin{figure}
   \centerline{\includegraphics[width=0.49\linewidth]{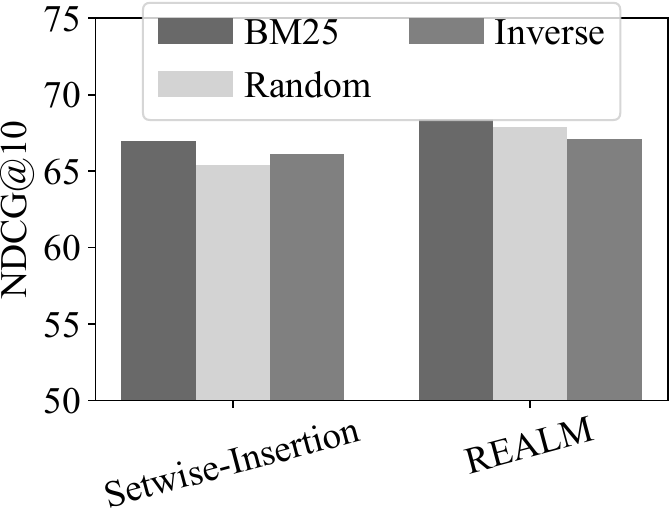}
   \includegraphics[width=0.49\linewidth]{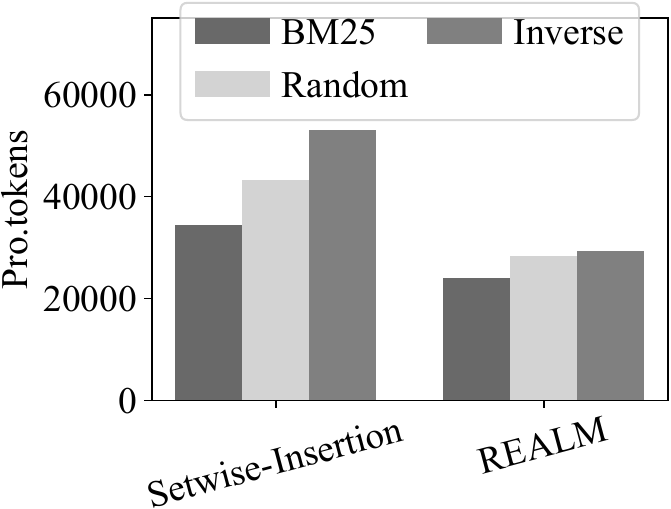}}
    \caption{Sensitivity to initial ranking order on TREC DL 2020 using Flan-T5-XL.}
    \label{fig:sensitivity}
\end{figure}

\stitle{Supplementary evaluation on BEIR datasets.} We further evaluate all methods on four BEIR datasets: Covid, SciFact, DBpedia, and NFCorpus, using Flan-T5-Large model; the corresponding results are reported in Table~\ref{tab:supple}. These results are consistent with our claims: {\llmrank} outperforms all existing approaches. In particular, {\llmrank} improves NDCG@10 by $2.6-27.6$, while reducing the number of inferences by $39.6-84.8\%$ and cutting inference time by $30.8-93.6\%$. In terms of prompt token usage, {\llmrank} lowers cost by $39.6-89.2\%$. Note that in NFCorpus, many queries have fewer candidate passages than in other datasets, which explains the shorter processing time.

\stitle{{\llmrank} vs Setwise-Insertion on the initial ranking.} 
Among existing baselines, Setwise-Insertion offers a reasonable balance between performance and efficiency, utilizing structured comparisons to reduce the number of LLM calls while maintaining competitive ranking quality. However, its effectiveness can still be affected by the quality of the initial document ordering.

Figure~\ref{fig:sensitivity} compares Setwise-Insertion and {\llmrank} on the TREC DL 2020 dataset under three initial document orderings: BM25, Inverse, and Random. The left plot reports NDCG@10, while the right plot shows the corresponding prompt token usage.
Here, Inverse refers to reversing the original BM25 ranking (i.e., least relevant documents placed first), while Random denotes a random permutation of the BM25-ranked list.

In terms of NDCG, the two methods perform comparably: Setwise-Insertion’s best and worst scores differ by 1.6 points, while {\llmrank} shows a smaller gap of 1.3 points, indicating slightly better stability.
However, the difference becomes more pronounced when comparing token efficiency. Because Setwise-Insertion relies more heavily on the assumptions of the initial ranking, it requires significantly more insertion operations when the initial order is suboptimal (e.g., under Inverse). This leads to substantially higher prompt token usage, whereas {\llmrank} is less affected by the quality of the initial ranking and maintains consistently low token consumption across all input orders.

\begin{table}
  \centering
    \resizebox{1\linewidth}{!}{ 
  \begin{tabular}{c|c|cccc}
    \hline
    \multirow{2}{*}{\textbf{LLM}}&\multirow{2}{*}{\textbf{Method}} & \multicolumn{4}{|c}{\textbf{Avg. Performance on TREC DL}} \\
    \cline{3-6}
    &&\textbf{N@10}&\textbf{\#Inf.}&\textbf{P. tks}&\textbf{Lat.(s)}\\\hline
     NA&BM25& 49.3  &- &- &-  \\\hline
     \hline
     \multirow{4}{*}{\rotatebox{90}{{\small Flan-T5-Large}}}&w/o modeling&62.0& 109.2&35064.2&5.4\\\cline{2-6}
     &w/o recursive&63.0&50.0&16089.0&2.6\\\cline{2-6}
     &w/o opt.&64.2&112.5&36335.6&5.6\\\cline{2-6}
     &{\llmrank}&\textbf{\underline{65.0}}&76.8&24375.3&3.8\\\hline
     \hline
 
     

     
    \multirow{4}{*}{\rotatebox{90}{{\small Flan-T5-XL}}}&w/o modeling&66.2&114.9&37003.4&5.8\\\cline{2-6}
    &w/o recursive&68.3&50.0&16089.0&2.6\\\cline{2-6}
    &w/o opt.&68.9&118.6&38409.6&6.0\\\cline{2-6}
    &{\llmrank}&\textbf{\underline{69.5}}&78.0&24928.3&3.9\\\hline
    \hline
    \multirow{4}{*}{\rotatebox{90}{{\small Flan-T5-XXL}}}&w/o modeling&68.7&113.7&36696.1&11.5\\\cline{2-6}
    &w/o recursive&68.6&50.0&16089.0&5.0
     \\\cline{2-6}
     &w/o opt.&68.9&119.2&38846.4&12.1 \\\cline{2-6}
     &{\llmrank}&\textbf{\underline{70.2}}&75.3&24209.7&7.4\\\hline
    \end{tabular}
}
    \caption{Ablation study.}
  \label{tab:ablation}
\end{table}

\subsection{Analysis}
\stitle{Ablation study.} Table~\ref{tab:ablation} presents the results of ablation study, comparing the full \textsc{\llmrank} system with three reduced variants by disabling key components: (1) \textsc{w/o modeling}, which removes uncertainty modeling and uses QuickSelect~\cite{hoare1961algorithm} to retrieve top-$k$; (2) \textsc{w/o recursive}, which disables recursive refinement; and (3) \textsc{w/o optimization}, omitting pivot optimization.

Removing any of these components leads to a consistent drop in performance. Disabling uncertainty modeling (\textsc{w/o modeling}) results in a $1.5-3.3$ decrease in NDCG@10 across all models, highlighting the value of Gaussian-based relevance modeling. The absence of recursive reasoning (\textsc{w/o recursive}) also causes noticeable degradation, underscoring the benefit of multi-round refinement. Lastly, disabling pivot-centric optimization (\textsc{w/o optimization}) nearly doubles latency-for example, from $7.4$s to $12.1$s with Flan-T5-XXL-confirming that our pivot selection and partitioning strategy substantially improves efficiency without compromising effectiveness.

\begin{figure}
   \centerline{\includegraphics[width=1\linewidth]{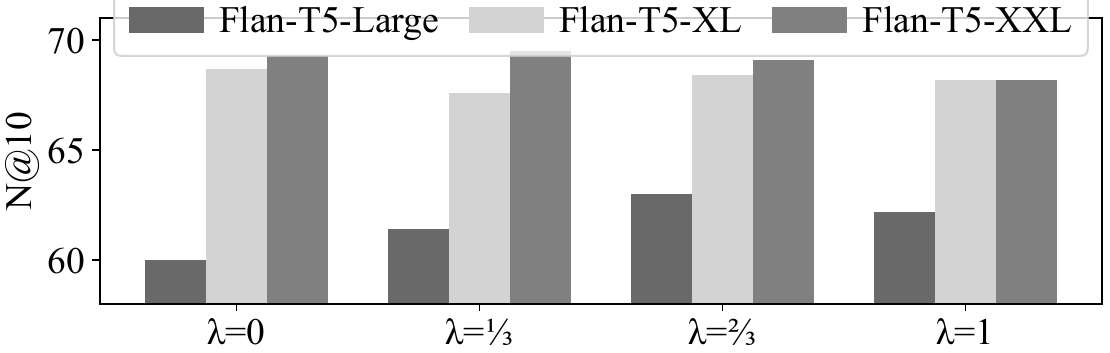}}
   \includegraphics[width=1\linewidth]{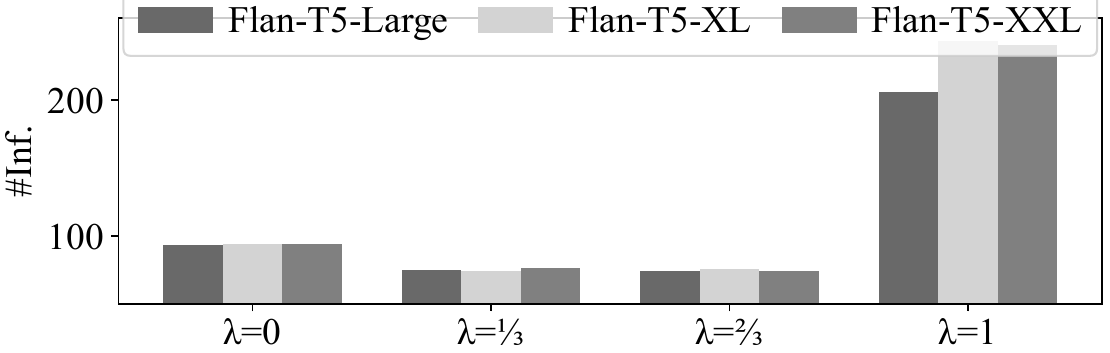}
    \caption{Analysis of Hyperparameter $\lambda$.}
    \label{fig:para}
\end{figure}

\stitle{Hyperparameter analysis.} 
We analyze the mixing coefficient $\lambda$ introduced in Section~\ref{subsec:quicksort}: specifically, we sweep $\lambda \in \{0,\tfrac{1}{3},\tfrac{2}{3},1\}$ and evaluate with Flan-T5-Large, XL, XXL models on TREC DL 2020 (see Fig.~\ref{fig:para}). The results reveal a clear cost-quality trade-off; overall, $\lambda=\tfrac{2}{3}$ yields the best performance-efficiency balance, which we adopt as the default setting in our main experiments.

\begin{table}
  \centering
  \resizebox{1\linewidth}{!}{ 
  \begin{tabular}{c|c|c|c|c}
    \hline
    \multirow{2}{*}{\textbf{LLM (Size)}}& \multicolumn{2}{|c|}{\textbf{TREC DL 2019}}& \multicolumn{2}{|c}{\textbf{TREC DL 2020}}\\
    \cline{2-5}
    &\textbf{N@10}&\textbf{lat.(s)}&\textbf{N@10}&\textbf{lat.(s)}\\\hline
     NA& 50.6   &-  &48.0 &- \\\hline
     Flan-T5-Large (770M)&67.0&3.9&63.0&3.6\\\hline
     Flan-T5-XL (3B)&70.5&4.0&68.4&3.7\\\hline
     LLaMA3 (8B)&49.6& 30.9&43.6&27.3\\\hline
     Flan-T5-XXL (11B)&71.2&7.5&69.1&7.3\\\hline
     Flan-UL2 (20B)&\textbf{\underline{72.2}}&13.2&\textbf{\underline{71.4}}&12.8\\\hline
     LLaMA3 (70B)&72.0&81.4&68.4&80.2 \\\hline

    \end{tabular}
    }
    \caption{Performance across different LLMs.}
  \label{tab:llms}
\end{table}
\stitle{Performance across different LLMs.}
As shown in Table~\ref{tab:llms}, our method benefits from stronger LLMs, achieving higher re-ranking performance and reduced prompt usage. For instance, Flan-UL2 achieves the best results on TREC DL datasets with an average NDCG@10 of $71.8$, while Flan-T5-XXL and Flan-T5-XL reach $70.2$ and $69.5$, respectively, surpassing the no-LLM baseline (48.0) and LLaMA models by a large margin.

In contrast, decoder-only architectures such as LLaMA-8B yield substantially lower performance (e.g., $49.6$ on TREC DL 2019), despite incurring $30.9$s per query. This discrepancy is partly due to their limited formatting capabilities and strong positional bias. In particular, we observe that LLaMA-8B selects the first option in approximately $85.6\%$ of cases on our balanced binary-choice tasks constructed from the TREC DL 2019 dataset (see Appendix~\ref{sec:cap}), indicating a strong positional bias that undermines its effectiveness for re-ranking.





\section{Conclusion}
We present {\llmrank}, an uncertainty-aware re-ranking framework. By modeling document relevance as Gaussian distributions and refining them through recursive comparisons and Bayesian aggregation, {\llmrank} achieves both high effectiveness and efficiency. Experiments across multiple LLMs and TREC benchmarks demonstrate that {\llmrank} consistently outperforms existing re-ranking methods across various configurations.  


\section*{Limitations}
Our current design choices are partially constrained by resource and space limitations.
While our relevance modeling is, in principle, compatible with a broad range of re-ranking methods, we center our evaluation on the specific framework we designed that achieved the strongest empirical performance. 
Besides, we have not considered skew-normal distributions for modeling relevance; while they may better capture the skewed score distributions given by the retriever, they would introduce additional complexity and inference costs. We leave the theoretical and empirical investigation of it to future work.
Last, we conduct experiments using open-source models, including the Flan-T5 and LLaMA 3 series. We did not include closed-source models such as GPT-3.5 or GPT-4 accessed via API. Future extensions may consider such models to provide a more complete empirical picture of {\llmrank}.


\section*{Acknowledgments}
We appreciate the anonymous reviewers for their constructive comments that helped refine and expand this work.
This work was in part supported by the NSF CAREER Award No. 2326141, and NSF Awards 2212370, 2319880, 2328948, 2411294, and 2417750.
This work was also performed under the auspices of the U.S. Department of Energy by Lawrence Livermore National Laboratory under Contract No. DE-AC52-07NA27344, and was supported by the U.S. Department of Energy, Office of Science, Advanced Scientific Computing Research (ASCR) program, under SC-21. LLNL-CONF-2010514.
This research used resources of the Oak Ridge Leadership Computing Facility at the Oak Ridge National Laboratory, which is supported by the Office of Science of the U.S. Department of Energy under Contract No. DE-AC05-00OR22725. 
%
The United States Government retains, and the publisher, by accepting the article for publication, acknowledges that the United States Government retains a non-exclusive, paid-up, irrevocable, world-wide license to publish or reproduce the published form of this manuscript, or allow others to do so, for United States Government purposes.



\bibliography{anthology,custom}

\appendix
\section{Bayesian Rating Systems}
\label{app:rating}
Bayesian rating systems provide probabilistic frameworks to estimate latent skill levels or quality scores of entities based on observed outcomes of comparisons or matches. Such systems leverage Bayesian inference principles, combining prior knowledge with observed data to update skill estimations dynamically. The general characteristics include modeling uncertainty explicitly, supporting incremental updating, and providing robustness to noise and incomplete data. Two widely-adopted Bayesian rating systems are Elo~\cite{elo} and its more advanced successor, TrueSkill~\cite{herbrich2006trueskill}, which progressively extend rating complexity and flexibility.

\subsection{Elo Rating System}
The Elo rating system is a foundational Bayesian rating method originally designed to quantify the relative skill levels of chess players. In this system, each player's ability is represented by a single numerical rating. When two players compete, the ratings are updated based on the observed outcome compared to the expected outcome calculated from current ratings. A player's rating increases after wins against higher-rated opponents and decreases upon losses or unexpected outcomes. The Elo system's simplicity and adaptability make it particularly effective in scenarios involving sequential pairwise competitions.

\subsection{TrueSkill Rating System}
TrueSkill, introduced by Microsoft, generalizes the Elo rating system by explicitly modeling player skills using probability distributions rather than single scalar values. Specifically, TrueSkill represents each player's skill as a Gaussian distribution characterized by two parameters: a mean $\mu$ reflecting the estimated skill level, and a standard deviation $\sigma$ capturing the uncertainty of this estimate. Following each match, TrueSkill applies approximate Bayesian inference to update these parameters according to the observed results, factoring in the certainty of each player's current rating. This mechanism enables TrueSkill to naturally handle multiplayer and team-based matches, uncertain outcomes, and noisy comparisons.

\subsection{Bayesian Update Details}
\label{app:update}

We adopt a Gaussian-based update rule derived from the 1v1 setting in TrueSkill~\cite{herbrich2006trueskill}, adapted for document ranking.

Given two documents \( D_i \) and \( D_j \) with current relevance estimates \( \mu_{i,0}, \sigma_{i,0}^2 \) and \( \mu_{j,0}, \sigma_{j,0}^2 \), we define the following intermediate quantities:

\[
\delta = \mu_{i,0} - \mu_{j,0}, \quad
c^2 = \sigma_{i,0}^2 + \sigma_{j,0}^2 + 2\beta^2,
\]
\[
t = \frac{\delta}{\sqrt{c^2}}, \quad
v(t) = \frac{\phi(t)}{\Phi(t)}, \quad
w(t) = v(t)(v(t) + t),
\]

where \( \phi(t) \) and \( \Phi(t) \) denote the probability density function and cumulative distribution function of the standard normal distribution, respectively. The parameter \( \beta \) is a fixed constant that controls comparison noise; we follow the TrueSkill default and set \( \beta = \mu_0 / 3 \).

We then define the Bayesian updates to the relevance distribution of \( D_i \), under two possible outcomes:

\paragraph{If \( D_i \) wins over \( D_j \):}
\[
\mathcal{N}(\mu_{\text{win}}, \sigma^2_{\text{win}})
\]
\begin{align*}
\Delta\lambda^{+} &= \frac{\sigma_{i,0}^4}{c^2} \cdot w(t), \\
\Delta\tau^{+} &= \frac{\sigma_{i,0}^2}{\sqrt{c^2}} \cdot v(t) + \mu_{i,0} \cdot \Delta\lambda^{+}, \\
\lambda_{\text{win}} &= \lambda_{i,0} + \Delta\lambda^{+}, \\
\tau_{\text{win}} &= \tau_{i,0} + \Delta\tau^{+}, \\
\mu_{\text{win}} &= \frac{\tau_{\text{win}}}{\lambda_{\text{win}}}, \quad
\sigma^2_{\text{win}} = \frac{1}{\lambda_{\text{win}}}.
\end{align*}

\paragraph{If \( D_i \) loses to \( D_j \):}
\[
\mathcal{N}(\mu_{\text{loss}}, \sigma^2_{\text{loss}})
\]
\begin{align*}
\Delta\lambda^{-} &= \frac{\sigma_{i,0}^4}{c^2} \cdot w(-t), \\
\Delta\tau^{-} &= -\frac{\sigma_{i,0}^2}{\sqrt{c^2}} \cdot v(-t) + \mu_{i,0} \cdot \Delta\lambda^{-}, \\
\lambda_{\text{loss}} &= \lambda_{i,0} + \Delta\lambda^{-}, \\
\tau_{\text{loss}} &= \tau_{i,0} + \Delta\tau^{-}, \\
\mu_{\text{loss}} &= \frac{\tau_{\text{loss}}}{\lambda_{\text{loss}}}, \quad
\sigma^2_{\text{loss}} = \frac{1}{\lambda_{\text{loss}}}.
\end{align*}

\section{Implementation Details}
\label{app:imp}
\subsection{Detailed Explanation of Datasets}

The TREC Deep Learning (DL) 2019~\cite{craswell2020overview} and 2020~\cite{craswell2021overview} datasets are benchmark collections designed to evaluate document ranking systems in complex information retrieval tasks. Both datasets are based on queries derived from real-world search logs and are built on top of the MS MARCO~\cite{nguyen2016ms} passage and document corpora. The TREC DL 2019 dataset includes $43$ queries with graded relevance judgments, while the 2020 version expands the test set to $54$ queries. All documents are written in English and drawn from a large web-scale corpus, with each query typically associated with hundreds to thousands of candidate passages.

We also evaluate on four datasets from the BEIR benchmark~\cite{BEIR}: Covid, SciFact, DBPedia, and NFCorpus. These collections complement TREC DL by covering diverse domains such as biomedical search, scientific fact verification, entity-centric retrieval, and consumer health information, thereby broadening the evaluation scope beyond general web search.

Together, these datasets emphasize fine-grained relevance estimation and are widely adopted for benchmarking re-ranking methods. For all experiments, the reported NDCG@10 scores are averaged over the entire dataset with a single run, ensuring stable and reliable evaluation results.

\subsection{Parameter Settings}
For a fair comparison, we set the number of comparison rounds for TourRank and PRP-Graph to $10$. Since the original implementation of TourRank only supports the OpenAI API, we re-implemented it with a T5-based interface. For Setwise-Insertion, we adopt the best-performing variant, \textit{Setwise Insertion Sort Compare Prior}, as reported in their paper~\cite{podolak2025beyond}. All other baselines are used with their default hyperparameters. In our method, we set $\lambda = 2/3$ (see Section~\ref{subsec:quicksort}) to balance effectiveness and efficiency.

\subsection{Prompts}
For TourRank, we adopt their default listwise prompt: 

\begin{tcolorbox}[colback=gray!10!white, colframe=black, boxrule=0.6pt, arc=3pt, width=\linewidth]
\textbf{System Prompt}

You are an intelligent assistant that can compare multiple documents based on their relevance to the given query.
\end{tcolorbox}

\begin{table*}
  \centering
    \resizebox{1\linewidth}{!}{ 
  \begin{tabular}{c|ccccc|ccccc}
    \hline
    \multirow{2}{*}{\textbf{LLM}}& \multicolumn{5}{|c|}{\textbf{TREC DL 2019}}& \multicolumn{5}{|c}{\textbf{TREC DL 2020}}\\
    \cline{2-11}
    &\textbf{N@10}&\textbf{\#Inf.}&\textbf{P. tks.}&\textbf{G. tks.}&\textbf{Lat.(s)}&\textbf{N@10}&\textbf{\#Inf.}&\textbf{P. tks.}&\textbf{G. tks.}&\textbf{Lat.(s)}\\\hline

     Flan-T5-Large
     &50.9&	9&	4608&	61.1&	2.2	&48.4	&9&	4608&65.0&2.3\\\cline{1-11}
     
     Flan-T5-XL&49.5&	9&	4608&80.9&4.1&47.0&9	&4608&79.4&4.1
     
     \\\cline{1-11}
      
Flan-T5-XXL	&50.5&9&9216&69.6&15.0&46.7&9&9216&60.9&14.5\\\hline      
 
    \end{tabular}
    }
    \caption{Additional Evaluation on RankGPT.}
  \label{tab:rankgpt}
\end{table*}

\begin{tcolorbox}[colback=gray!10!white, colframe=black, boxrule=0.6pt, arc=3pt, width=\linewidth]
\textbf{User Prompt}

I will provide you with the given query and \texttt{\{N\}} documents.

Consider the content of all the documents comprehensively and select the \texttt{\{M\}} documents that are most relevant to the given query: \texttt{\{query\}}.

The query is: \texttt{\{query\}}.

Now, you must output the top \texttt{\{M\}} documents that are most relevant to the query using the following format strictly, and nothing else.

Do not provide any explanation or commentary. Output format:

\texttt{Document 3, ..., Document 1}
\end{tcolorbox}

For PRP-Graph, we also adopt their default pairwise prompt:

\begin{tcolorbox}[colback=gray!10!white, colframe=black, boxrule=0.6pt, arc=3pt, width=\linewidth]
Given a query \{\texttt{query}\}, which of the following two passages is more relevant to the query?

\vspace{0.5em}
Passage A: \{\texttt{document\_1}\}

Passage B: \{\texttt{document\_2}\}

\vspace{0.5em}
Output Passage A or Passage B:
\end{tcolorbox}

For the remaining methods ({\llmrank}, Setwise-Heapsort, and Setwise-Insertion), we use a consistent setwise prompt of the following form:

\vspace{0.5em}
\begin{tcolorbox}[colback=gray!10!white, colframe=black, boxrule=0.6pt, arc=3pt, width=\linewidth]

Given a query \{\texttt{query}\}, which of the following passages is the most relevant to the query?

\{\texttt{passages}\}

Output only the passage label of the most relevant passage:
\end{tcolorbox}

\vspace{0.5em}
For Setwise-Insertion, we additionally append the following sentence to the prompt:

\vspace{0.5em}
\begin{tcolorbox}[colback=gray!10!white, colframe=black, boxrule=0.6pt, arc=3pt, width=\linewidth]

If their relevance is similar, or none of them is relevant, output A.
\end{tcolorbox}

\vspace{0.5em}
This modification follows their original paper, which claims this change as a key contribution.

\subsection{Code Availability}

The source code of {\llmrank} is publicly available at:  
\url{https://github.com/Joeyw02/REALM}.



\section{Supplementary Evaluations}

\subsection{Additional Evaluation on RankGPT}

We additionally include RankGPT~\cite{sun-etal-2023-chatgpt} in the overall evaluation (Table~\ref{tab:rankgpt}). 
The results show that---just like TourRank---this list-wise approach 
is bounded by the base model's capacity: when the model cannot reliably 
handle long contexts, it struggles to produce a complete, well-ordered 
ranking, often generating partial or seemingly random sequences.
\subsection{Model Capability Analysis}
\label{sec:cap}

We evaluate the pairwise comparison capability of different LLMs by randomly sampling 500 document pairs per query from the TREC DL 2019 dataset. Each pair is presented to the model for binary relevance judgment. As shown in Table~\ref{tab:pair}, we observe that LLaMA 3 8B exhibits a strong position bias, often favoring the document appearing in a particular position regardless of content. In contrast, Flan-T5 models demonstrate more reliable behavior and stronger alignment with ground-truth preferences in pairwise comparisons.

\begin{table}
  \centering
  \begin{tabular}{|c|c|c|c|}
    \hline
    \textbf{Choice} & \textbf{Correct}&\textbf{Wrong}& \textbf{Accuracy}\\
    \hline
    \multicolumn{4}{|c|}{\textbf{LLaMA3 8B}}\\ \hline
    A (85.6\%)& 10258 & 8145  &55.7\% \\\hline 
    B (14.4\%) &  2669  &428   &86.2\%\\\hline 
    Total  & 12927 &8573& 60.1\%\\\hline

    \multicolumn{4}{|c|}{\textbf{LLaMA3 70B}}\\ \hline
    A  (54.9\%) &  9979  &  1815&84.6\%\\\hline 
    B    (45.1\%)&   8794   & 912 &90.6\%\\\hline 
    Total  & 18773&2727&87.3\%\\\hline

    \multicolumn{4}{|c|}{\textbf{FLAN-T5-XXL 11B}}\\ \hline
    A (43.9\%) &  8709 & 722 &92.3\% \\\hline 
    B (56.1\%)  & 10047  & 2022 &83.2\%\\\hline 
   Total  &  18756 & 2744&87.2\%  \\\hline

    \multicolumn{4}{|c|}{\textbf{FLAN-UL2 20B}}\\ \hline
    A (50.3\%) &  9684  & 1137  &89.5\% \\\hline 
    B (49.7\%)  &  9542  & 1137& 89.4\%\\\hline 
    Total  & 19226 &2274&89.4\%  \\\hline

    \end{tabular}
    \caption{Statistical summary of different models' choices in pair-wise comparison on TREC DL 2019.}
  \label{tab:pair}
\end{table}


\subsection{Iteration Analysis}

For the results reported in Table~\ref{tab:overall}, we further measured the average number of iterations required during inference on the TREC DL 2019 and 2020 datasets in Table~\ref{tab:iter}. Across different settings, the model required an average of $4.89$ rounds of iterations. This observation suggests that, given sufficient computational resources, our method can be naturally parallelized and thus remains efficient even with multiple iterative steps.

\begin{table}
  \centering
  \begin{tabular}{|c|c|c|}
    \hline
    \textbf{\small{Model}} & \textbf{\small{TREC DL 2019}}&\textbf{\small{ TREC DL 2020}}\\
    \hline
    \small{Flan-T5-Large}& 5.12 & 4.67  \\\hline 
    \small{Flan-T5-XL} &  4.98  &4.98 \\\hline 
    \small{Flan-T5-XXL}& 5.02&4.57\\\hline

    \end{tabular}
    \caption{Average number of iterations required
during re-ranking on the TREC DL 2019 and 2020
datasets.}
  \label{tab:iter}
\end{table}



\end{document}